\documentclass[fleqn,usenatbib]{mnras}

\usepackage[T1]{fontenc}
\usepackage{ae,aecompl}

\usepackage{graphicx}	
\usepackage{amsmath}	
\usepackage{amssymb}	
\usepackage{threeparttable}
\usepackage{multirow}

\newcommand{\ts}{\thinspace}
\newcommand{\gr}{$^{\circ}$}
\newcommand{\x}{$\times$}
\newcommand{\fl}{cm$^{-2}$s$^{-1}$keV$^{-1}$}

\title[LECX: a cubesat experiment to detect and localize cosmic explosions in hard X rays]{LECX: a cubesat experiment to detect and localize cosmic explosions in hard X rays}
\author[J. Braga et al.]{
 J.\ Braga,$^{1}$\thanks{E-mail: joao.braga@inpe.br}
 O.\ S.\ C.\ Dur\~ao,$^2$
 M. Castro,$^{1}$
 F.\ D'Amico,$^{1}$
 P.\ E.\ Stecchini,$^{1}$
 S.\ Amir\'abile,$^{1}$\newauthor
 F.\ Gonzalez Blanco,$^{1}$
 C.\ Strauss,$^{1}$
 W.\ Silva,$^{3}$
 V.\ R.\ Schad,$^{4}$
 and L.\ A.\ Reitano$^{1}$
\\
$^{1}$Instituto Nacional de Pesquisas Espaciais, Av.\ dos Astronautas 1758, S\~ao Jos\'e dos Campos, SP, Brazil\\
$^{2}$CRON Sistemas e Tecnologia Ltda., Av.\ Andr\^omeda, 693, sl.\ 708, S\~ao Jos\'e dos Campos, SP, Brazil\\  
$^{3}$Atlas Software e Solu\c{c}\~oes Ltda., Rua Dan\'ubio 217, Uberl\^andia, MG, Brazil\\  
$^{4}$Horus Eye Tech Engenharia de Sistemas, R.\ Prof.\ Roberval Fr\'oes, 390, sl.\ 202, S\~ao Jos\'e dos Campos, SP, Brasil\\ 
}

\date{Accepted XXX. Received YYY; in original form ZZZ}

\pubyear{2019}

\graphicspath{{./figs/}}

\begin{document}

\label{firstpage}
\pagerange{\pageref{firstpage}--\pageref{lastpage}}
\maketitle

\begin{abstract}

With the advent of the nanosat/cubesat revolution, new opportunities have appeared to develop and launch small ($\sim$\ts 1000 cm$^3$), low-cost ($\sim$\ts US\$ 1M) experiments in space in very short timeframes ($\sim$ 2\ts years). In the field of high-energy astrophysics, in particular, it is a considerable challenge to design instruments with compelling science and competitive capabilities that can fit in very small satellite buses such as a cubesat platform, and operate them with very limited resources. Here we describe a hard X-ray (30--200\ts keV) experiment, LECX (``Localizador de Explos\~oes C\'osmicas de Raios X'' -- Locator of X-Ray Cosmic Explosions), that is capable of detecting and localizing within a few degrees events like Gamma-Ray Bursts and other explosive phenomena in a 2U-cubesat platform, at a rate of $\sim${\bf 5 events year$^{-1}$.} In the current gravitational wave era of astronomy, a constellation or swarm of small spacecraft carrying instruments such as LECX can be a very cost-effective way to search for electromagnetic counterparts of gravitational wave events produced by the coalescence of compact objects.

\end{abstract}




\section{Introduction}

The design of competitive space instruments to detect X- and gamma-ray fluxes from astrophysical sources has always been a challenge due to several limitations, especially on weight, size and power consumption. In order to achieve good sensitivities for the desired instrumental parameters such as energy range, imaging capabilities, energy resolution and field-of-view (FOV), instrument builders have to carefully select detector types and structural parts that meet the constraints imposed by the available or envisaged resource possibilities. 

Normally, non-focusing instruments such as coded-aperture telescopes (see \citealp{2020PASP..132a2001B} for a recent review) need both large detector areas to maximize source {\bf count rates} and massive shielding systems to minimize background levels. With the current explosive growth of the nanosat phenomenon based especially on the so-called CubeSat platform (see \citealp{Thyrso2019} and references therein), new opportunities for fast-development, low-cost small instruments have appeared. 

In principle, X- and gamma-ray astrophysical instruments on such small satellites cannot compete with full-sized instruments operating on conventional large and complex satellite buses.  Nevertheless, for very specific science goals, it is possible to design instruments compatible with cubesat buses that can meet the desired requirements. In this paper we describe an instrument developed for a cubesat platform that is capable of detecting and locating relatively strong cosmic explosions that manifest themselves electromagnetically mainly in the hard X-ray/low energy gamma-ray range. The experiment, called ``Localizador de Explos\~oes C\'osmicas de Raios X'' (LECX -- Portuguese for Locator of X-ray Cosmic Explosions), will be sensitive enough to detect and localize within a few degrees {\bf events like} the well-known gamma-ray bursts (GRBs -- see \citealp{2018pgrb.book.....Z}) in the 30--200\ts keV energy range. With its 53\gr \x 53\gr FWHM (``Full Width at Half Maximum'') FOV, it is estimated that LECX will detect $\sim$\ts 5 GRBs per year.

In the recently-inaugurated mutimessenger astrophysics era, it is of paramount importance that wide-field space instruments constantly patrol the sky in order to instantly detect electromagnetic (EM) counterparts of gravitational wave (GW) and/or neutrino cosmic bursting events. With the increased sensitivities of ground-based observatories of gravitational waves (e.g.\ LIGO/VIRGO) and neutrinos (e.g.\ IceCube), it is expected that the rate of such events will gradually increase over the years. X- and gamma-ray space experiments will then have higher probabilities of contributing to multimessenger detections of such phenomena.

The technology being developed for LECX builds upon what has been developed for the protoMIRAX project \citep{Braga2015}, a balloon experiment that represents a prototype of the MIRAX ({\it Monitor e Imageador de Raios X}, in Portuguese) space mission \citep{2006AIPC..840....3B, 2004AdSpR..34.2657B}. 

LECX is in the integration and testing phase of the engineering model and constitutes the payload of the CRON-1/nanoMIRAX satellite \citep{Durao2019}, which is based on a 2U-cubesat platform. nanoMIRAX is being developed by the Brazilian private company CRON in partnership with INPE (National Institute for Space Research). In section \ref{sec:det} we describe the components of the detector system and associated electronics of the LECX instrument. In section \ref{sec:posit} we present the algorithm that we have developed to determine the sky positions of cosmic explosions that occur in the instrument FOV. In section \ref{sec:sens} we discuss the expected background that the instrument  will be subject to in low-Earth circular equatorial orbit and the expected sensitivity of the instrument for the detection of bright, short-lived point sources. In section \ref{sec:grb} we describe simulations of GRB detections with LECX {\bf that show the potential of the instrument to localize these sources}. In section \ref{sec:sat} we describe the satellite bus and its capabilities. Finally, in section \ref{sec:conc} we present our conclusions.

\section{The LECX instrument}
\label{sec:lecx}

\subsection{The detector system}
\label{sec:det}

\begin{figure}
\begin{center}
\includegraphics[width=0.5\hsize]{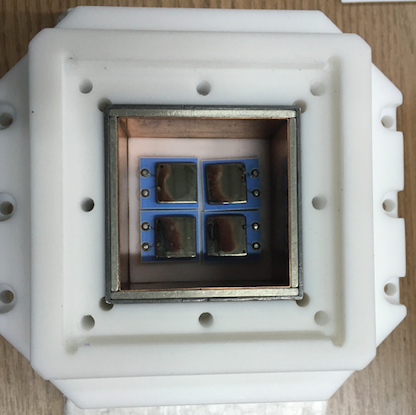}
\caption{Photograph of the detector plane of LECX at the bottom of the shielding box.}
\label{fig:detectors}
\end{center}
\end{figure}

LECX employs four CdZnTe (CZT) planar detectors in a 2x2 configuration. Each detector is a 10mm x 10mm square with a thickness of 2mm, operating from 30 to 200 keV, described in detail in \citet{Braga2015}. The lower limit is imposed by electronics noise and the higher limit is due to detector thickness. The separation between adjacent detector volumes is 3\ts mm due to the mechanical mounting and the presence of the detectors' ceramic (alumina) substrates. Figure \ref{fig:detectors} shows the detector plane. CZT room-temperature semiconductor detectors have been extensively used in astronomical X- and $\gamma$-ray space instruments due to their very high photoelectrical efficiency up to hundreds of keV and good energy resolution. They are also easy to handle and can be tiled to cover large surfaces. Pixelated CZT detectors can also be built for imaging instruments \citep{2018SPIE10762E..14P, 2013JInst...8P9010R}.

\begin{figure}
 \centering
\includegraphics[width=0.8\hsize]{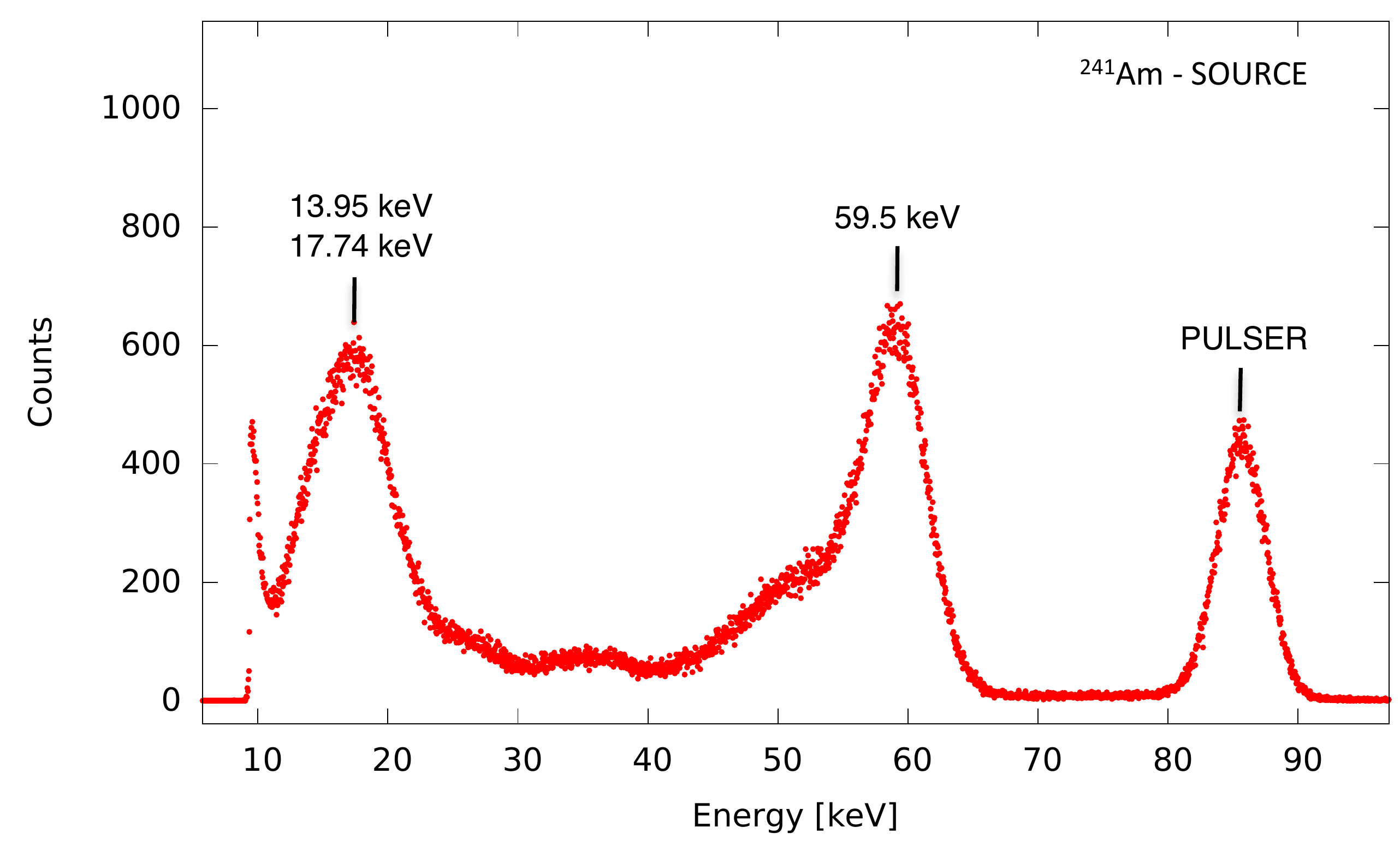}
\caption{Energy spectrum of one CZT detector exposed to a $^{241}$\!Am radioactive source. The peak on the right is produced by a fixed-height pulser for peak width comparison. Taken from \citet{Braga2015}.}
\label{fig:spectrum}
\end{figure}

Figure \ref{fig:spectrum} shows a typical energy spectrum of one detector, in this case with a 60\ts keV nuclear emission line coming from a $^{241}$\!Am radioactive source. The energy resolution is about 11 per cent {\bf at 60\ts keV}. The red wing at the 60\ts keV line is due to incomplete charge collection inside the detector material, an ubiquitous feature of CZT detectors.

The array is surrounded by a Pb (1.0mm), Sn (1.7mm) and Cu (0.3mm) graded shield box to minimize background and define the instrument FOV. The distance from the detector plane surface to the top of the shielding box is 20mm. In the upper part of the box there is an aperture of 23mm x 23mm that matches the detector plane area below (considering the gaps between the detectors). The area of the aperture is closed with a 0.4mm-thick carbon fibre plate to prevent the entrance of environment light, which induces noise in the detector electronics.

\begin{figure}
\begin{center}
\includegraphics[width=0.4\hsize]{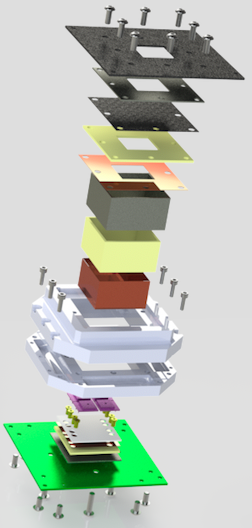}
\caption{Exploded view of the LECX detector system. From top to bottom, the carbon fibre sheets, the shielding layers and box walls (Pb-Sn-Cu), the teflon structure (2 parts), the detector plane, the bottom shielding layers and the cubesat PCB.}
\label{fig:exploded}
\end{center}
\end{figure}

\begin{figure}
\begin{center}
\includegraphics[width=0.45\hsize]{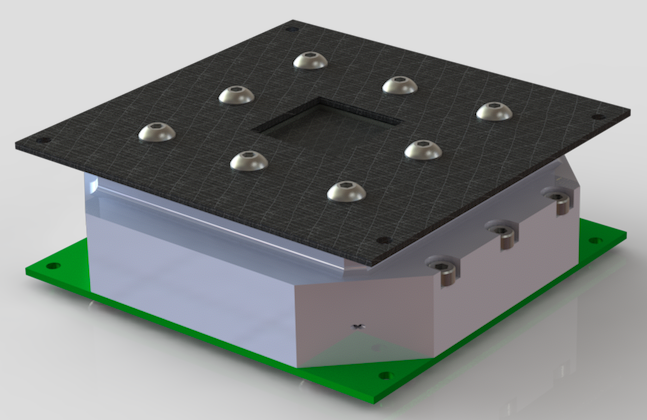}
\includegraphics[width=0.45\hsize]{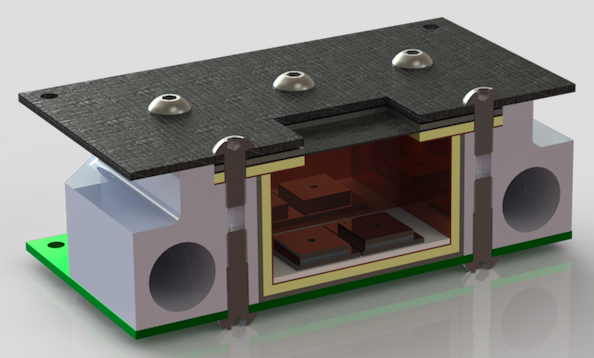}
\caption{An external view and a cutaway view of the LECX detector system in half, showing two of the four detectors (which are reflected on the internal copper plate). The upper carbon fibre plate is placed for structural purposes, whereas the bottom green plate is the DPCB (see text). We can see the teflon structure with the cylindrical space for the batteries, the passive shielding layers (Pb-Sn-Cu), the teflon plate underneath the detectors and the 0.4mm carbon fibre plate at the aperture entrance.}
\label{fig:half}
\end{center}
\end{figure}

The detector system is surrounded by a dielectric material (teflon) structure that provides mechanical support and housing for batteries and electronic parts. Teflon is widely used in space application due to its suitable mechanical properties. The whole system is mounted on a standard 89mm$\times$89mm cubesat printed-circuit board (the DPCB, for Detector Printed Circuit Board) at the top of the satellite structure (the ``top" direction hereafter refers to the instrument axis, i.e.\ the direction corresponding to the centre of its FOV). The front-end analog electronics for the detectors, which comprises four sets of charge pre-amplifiers and low-noise shaping amplifiers, is mounted on the opposite (bottom) side of the DPCB. This PCB, as well as the other two, has multiple interconnected copper layers to provide electrical shielding. At the bottom part of the detector substrates lies the bottom part of the shielding box, so that the two electric leads (the one that polarizes the anode with the reverse bias and the one carrying the charge signal pulse from the cathode) perforate the shielding material, in teflon tubes, in order to reach the PCB underneath. The radiation shielding is also connected to the common ground to contribute to the electrical shielding. Figure \ref{fig:exploded} shows an exploded-view diagram of the detector system mounted on this PCB and Figure \ref{fig:half} shows two different views of the system. 

The performance of the detector plane has been demonstrated in the protoMIRAX experiment \citep{Braga2015}. The reverse bias of $-200$\ts V was carefully chosen so as to minimize signal losses due to incomplete charge collection within the CZT material (manifested by the red wing of the 60\ts keV photopeak in Figure \ref{fig:spectrum}) whilst keeping very low levels of dark current. The long-duration CR1216 Lithium Manganese Dioxide batteries used for the detector power supply are placed close to the detector plane and encapsulated in the teflon structure to avoid current leakage.

A second PCB (the TPCB, for Time Printed-Circuit Board), mounted underneath the DPCB, houses the four Height-Time Converter (HTC) electronics, which linearly converts the heights of the pulses coming from the detectors to a digital high level signal. A third PCB, the experiment on-board-computer (EOBC), houses the digital electronics, which is responsible for performing the following tasks: (a) receive signals from the 4 outputs of the TPCB; (b) measure the time duration of the correspondent high level signals and convert them into digital 8-bit words (this is proportional to the deposited energy of each event, divided in 256 channels), (c) flag the individual detector were the event occurred and convert it into a 2-bit word; (d) tag each event with the Universal Time from the satellite on-board computer (OBC), {\bf that uses a GPS receiver, with a resolution of 255$\mu$s}; (e) build the event packages with time, detector ID and energy information (each event will generate a 24-bit word); (f) store data files and send copies to the spacecraft OBC for transmission to ground. The EOBC uses a commercial low power PIC24F32 microcontroller very suitable for this specific application. The power consumption of the LECX electronic system is less than 800\ts mW.

According to simulations of the radiation level to be measured in orbit (see section \ref{sec:sens}) the LECX payload will produce $\sim$\ts 100 bits/s in nominal operation, which will generate $\sim$\ts 540 kbits/orbit and $\sim$\ts 8.6 Mbits/day of data. Assuming at least one 10-minute ground station passage every orbit, this will require $\sim$\ts 900 bits/s telemetry capacity, well below the envisaged capabilities of the satellite.

The three PCBs of the nanoMIRAX payload, which comprise the LECX experiment, fill the first ``U'' of the satellite and constitute the payload module. The second ``U'', the service module, mounted on the bottom of the payload module, houses the satellite subsystems described in section \ref{sec:sat}. 

\subsection{The source position determination algorithm}
\label{sec:posit}

With the four CZT detectors placed inside the passive shielding box, as described in the last section, the FOV of the detector system is a square region of the sky of $53^\circ\times 53^\circ$ FWHM ($\approx$ 7 per cent of the sky) and $90^\circ\times 90^\circ$ FWZI (``Full Width at Zero Intensity''). In this section we describe a new algorithm we have devised to determine the celestial coordinates of strong cosmic explosions based on the intensities measured in each detector during the event. 

In standard collimated high-energy detector systems, the fluxes of astrophysical sources are determined by subtracting a background level, measured separately,  from the amount of radiation measured when the source is in the collimator FOV. For pixelated detector planes, one can use a coded mask in the aperture to obtain an image of the source field within the FOV (see, e.g. \citealp{2020PASP..132a2001B}). In the LECX cubesat experiment described here, we have only 4 pixels (the four planar detectors) and a limited sensitivity due to the fact that the total area is only 4 cm$^2$. Since we are interested in detecting strong, short-duration point-source cosmic explosions one at a time, a coded mask placed in the experiment's aperture will not be adequate to use due to three reasons: (a) the results to be obtained are not images of source fields, but only measured fluxes of short-duration ($\sim$\ts seconds) point sources. The known persistent or transient astrophysical sources will be too weak to detect on those timescales; (b) the advantage of coded-mask systems in measuring source and background simultaneously are not important here, since we will have plenty of time to measure the background anyway, and in the same region of the sky; and (c) there are no ``URA-like" mask patterns with $2\times 2$ dimensions that could produce images with no intrinsic noise (see \citealp{1978ApOpt..17..337F}), i.e.\ there are no $2\times 2$ patterns with a 2D discrete Dirac-$\delta$ autocorrelation function. We decided to develop and use instead a different algorithm that takes into account the ``shadows'' projected on the detector plane by the walls of the shielding box. This has the additional advantage of not partially blocking the FOV with a mask. 

During the detection of a radiation burst coming from a single direction (representing a cosmic explosion), the $x$ and $y$ extensions (with respect to the square shielding box and detector orientations) of the wall shadows on the detector plane is a function of the azimuth and zenith angles of the incoming photons. Since the number of source counts in each detector is proportional to the illuminated area, we can determine the source position in the sky by measuring the counts in each detector when the count rates sudden increase during the occurrence of such an event in the FOV. 

\begin{figure}
 \centering
\includegraphics[width=0.4\hsize]{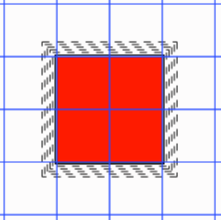}
\includegraphics[width=0.4\hsize]{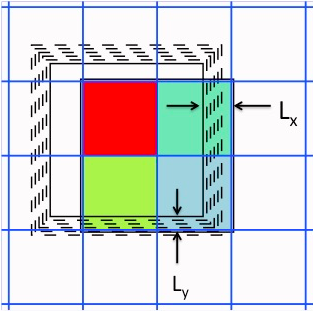}
\includegraphics[width=0.15\hsize]{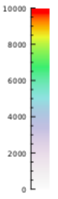}
\caption{The detector plane of LECX, viewed from above with a coordinate grid shown in blue. The illuminated area lies inside the black square with shaded regions outside. The counts are shown in a colour scale. {\bf Left:} incidence from the instrument axis (the ``zenith''); {\bf Right:} incidence from an azimuth of 333$^\circ$ (with respect to the instrument coordinate system) and a zenith angle of 13$^\circ$. The position determination algorithm reproduces exactly the incidence direction from the four count values.}
\label{fig:simu_1}
\end{figure}

To explain the method, let us first define a coordinate system with axes $x$ and $y$ along the detector sides for a given detector plane orientation in which North is to the top ($y$) and East is to the right ($x$); see Figure  \ref{fig:simu_1}. The relevant units of distance are the detector side $s$ (10mm in this case) and the height of the shielding walls above the detector surface, $H$ (20mm).  

The diagrams in Figure \ref{fig:simu_1} show simulations in which the 4 detectors are depicted along with a surrounding area. A black square, with dashed lines outside it, represents the illuminated area. (In these simulations the detector areas are exactly adjacent, not separated by 3\ts mm as in the real case, but this introduces  {\bf insignificant} differences in the results when statistics are taken into account.) On the left the source is at infinity along the instrument's axis, so the 4 detectors are equally illuminated. On the right, the source photons are coming from an azimuth ($A$) of 333$^\circ$ (4th quadrant) and a ``zenith'' angle ($z$) of 13$^\circ$, both with respect to the instrument coordinate system. The colour code gives the count scale in which 10k counts were initially thrown at each detector before correction by the shadows. The $x$ and $y$ extensions of the shadows ($L_x$ and $L_y$; see Figure \ref{fig:simu_1}) will clearly be
\begin{equation}
L_x = H \tan z \cos A\\ ; \\ L_y = H \tan z \sin A.
\label{eq:offsets}
\end{equation}
Now, starting from the upper left corner, let us call the detector illuminated areas $A_{11}, A_{12}, A_{22}$ and $A_{21}$, going clockwise. Then 
\begin{equation*} 
A_{11} = s^2 \;\;\;\;\;\;\;\;\;\;\;\;\;\;\; ; \;\;\;   A_{12} = s(s - L_x) 
\end{equation*}
\vspace{-0.8cm}
\begin{equation}
A_{21} = s(s - L_y)\;\;\; \; ;  \;\;\;  A_{22} = (s - L_x)(s - L_y)   
\label{eq:areas} 
\end{equation}

In this particular case, since detector $(1,1)$ has the highest number of counts, the azimuth angle clearly ranges from 270$^\circ$ to 360$^\circ$. Similar schemes, with equivalent equations, hold for the other three quadrants. In the general case, the algorithm first identifies the quadrant where the incoming direction lies by finding the detector with the most counts. Then it flags the other detectors in a decreasing order of counts. We then solve equations in the form of equations \ref{eq:areas} (depending on the quadrant) for $L_x$ and $L_y$, and finally use equations \ref{eq:offsets} to find $A$ and $z$. The algorithm reproduces the incoming angles exactly when no statistical fluctuations are taken into account.
 
\begin{figure}
 \centering
\includegraphics[width=0.9\hsize]{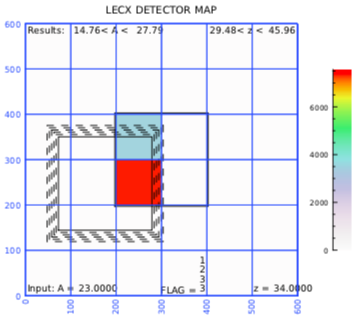}
\caption{Simulation with incidence from an azimuth of 23$^\circ$ and a ``zenith'' angle of 34$^\circ$. The position determination algorithm can only determine ranges for the incidence angles.}
\label{fig:grande_angulo}
\end{figure}

\begin{figure}
 \centering
\includegraphics[width=0.9\hsize]{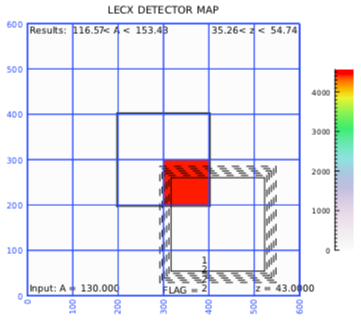}
\caption{Simulation with incidence from an azimuth of 130$^\circ$ and a ``zenith'' angle of 43$^\circ$.}
\label{fig:one_detector}
\end{figure}

For large zenith angles of incidence, $z > \arctan(s/H)$ in the orthogonal directions up to $z > \arctan(\sqrt{2}s/H)$ in the diagonal directions, the radiation coming from the source will miss one line of detectors entirely. In those cases, the algorithm can only determine ranges of $A$ and $z$ due to the lack of information from the four detector count numbers, which prevents us from calculating count ratios between detectors. As an example, in Figure \ref{fig:grande_angulo} we show a simulation in which the burst came from $A = 23^\circ$ and $z = 34^\circ$, hence outside the FWHM FOV. In this case, we can only determine that the position in the sky was $14^\circ\!\!.8 < A < 27^\circ\!\!.8 $ and $ 29^\circ\!\!.5 < z < 46^\circ\!\!.0 $. Another example is in Figure \ref{fig:one_detector}; here the source is at $A = 130^\circ$ and $z = 43^\circ$ so that only one detector is illuminated; all we can say by measuring the counts is that $ 116^\circ\!\!.6 < A < 153^\circ\!\!.4 $ and $ 35^\circ\!\!.3 < z < 54^\circ\!\!.7$.


Even in those cases, it is noteworthy that we can still get some localization despite using a very simple experimental setup. In particular, if the EM burst is coincident with a gravitational wave event, {\bf any independent localization may prove to be useful since the GW laser-interferometric detectors have poor localization capabilities}. With the FOV of LECX, it is expected that the mission will detect $\sim$\ts 5 cosmic explosions per year {\bf (see section \ref{sec:grb})}.

\subsection{Background and sensitivity}
\label{sec:sens}

\begin{figure}
 \centering
\includegraphics[width=0.9\hsize]{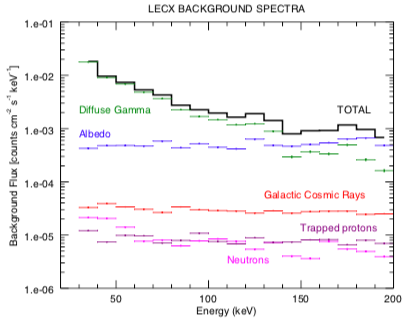}
\caption{Simulated spectra of the main components of the LECX background in equatorial LEO, considering the 4 detectors {\bf as one unit and a spacecraft attitude in which the instrument's axis is pointed to the zenith}.}
\label{fig:bkg}
\end{figure}

Estimation of the energy spectrum of the background signal and its spatial distribution over the detector plane is crucial for the design of hard X-ray and gamma-ray astronomy telescopes. In the case of an observation of a point source from an orbital space platform (i.e. a satellite), the background consists in the diffuse EM radiation coming through the telescope aperture, emission from other sources in the FOV, and the instrumental background, which arises from interaction of high-energy particles with the detectors and surrounding material \citep{Gehrels1985, Dean2003, Gruber1999, Ajello2008, Armstrong1973}. Therefore, in order to foresee the performance of LECX in orbit and its sensitivity to cosmic explosions, we {\bf need} to have a good estimate of the background radiation the detectors will measure. We have calculated this using the well-known GEANT4 package \citep{Agostinelli2003, Sarkar2011}. Details of our procedure to calculate the background of an instrument from  angle-dependent spectra of photons and particles in space can be found in \citet{2016MNRAS.459.3917C}. Considering  a near-equatorial low-Earth orbit (LEO) and a mass model of the LECX experiment, we have obtained the main components of the expected background in orbit, outside the {\bf South Atlantic Geomagnetic Anomaly (SAGA)}. This is shown in Figure \ref{fig:bkg}.

We can see that the main contributions come from the diffuse EM flux entering the aperture (up to $\sim$\ts 140\ts keV) and the albedo radiation coming from the Earth atmosphere (above $\sim$\ts 150\ts keV). {\bf During the occurrence of a cosmic explosion event, it is reasonable to consider that the Compton-scattered events and fluorescence on the collimator walls will not add significantly to the background that will be present during the event, since the graded-shield walls were specifically designed to minimize these radiations with the help of GEANT4 simulations (see \citealp{2016MNRAS.459.3917C}).} 

The sensitivity of LECX can be calculated considering that the number of counts in each detector, for a given integration time and a given energy bin, obey Poisson statistics. In the case of cosmic explosions observations, what will be measured are sudden increases in the total count rate that will last typically from {\bf a fraction of a second} to tens of seconds during nominal operation. A trigger mechanism will detect these surges and automatically put the experiment in burst mode, {\bf which will end when the low count rates resume. During burst mode, the satellite service module will provide more frequent attitude information data}. The electronics is designed so that all events will be time-tagged and stored {]\bf onboard} even if the count rate increases by a factor of $\sim$\ts 100. {\bf All science and spacecraft data will be transmitted to the ground during the ground station passages.}

Under nominal operations, the detector system will be measuring background radiation before and after the detected bursting events. We will select the best timescales to get good statistics for the background measurements. Since we may have significant background variations during one orbit, we also have to minimize this integration time to get more accurate values with respect to the background in effect during the occurrence of the event. During the event, we subtract the background counts from the total counts in each selected energy bin to obtain the source spectrum. It is easy to show that the minimum detectable flux, in photons cm$^{-2}$s$^{-1}$keV$^{-1}$, for each energy bin $\Delta E$ centred on energy $E$ will be
\begin{equation}
F_{\rm min} (\Delta E) = \frac{N_\sigma}{\epsilon(E)} \sqrt{\frac{B(E)}{A_{\rm det} \Delta E}\left[ \frac{1}{T_B} + \frac{1}{T_S} \right]}
\label{eq: sens}
\end{equation}
where $N_\sigma$ is the statistical significance (signal-to-noise ratio), $\epsilon(E)$ is the detector efficiency, $B(E)$ is the background in counts cm$^{-2}$s$^{-1}$keV$^{-1}$, $A_{\rm det}$ is the detector geometrical area in cm$^2$ (in this case 4\ts cm$^2$), $\Delta E$ the energy bin in keV, $T_B$ the time spent measuring background and $T_S$ in the time spent measuring source$+$background (both in seconds). 

\begin{figure}
 \centering
\includegraphics[width=0.9\hsize]{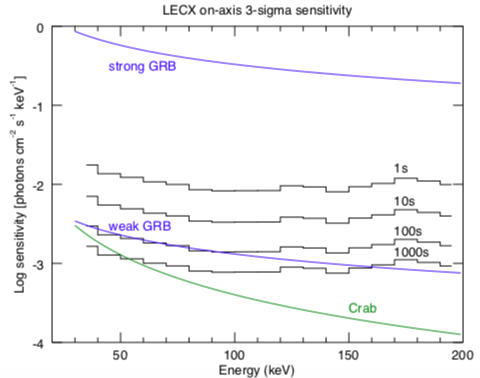}
\caption{{\bf On-axis} 3-$\sigma$ sensitivity of LECX for different on-source integration times, considering 10 minutes for background integration. Strong and weak GRB spectra are shown in blue and the Crab spectra is shown in green for comparison.}
\label{fig:sens}
\end{figure}

According to the expected count rates, 5 minutes is a reasonable time to measure background before and after a triggered event, considering a $\sim$\ts 90-minute orbit. Using then 10 minutes for background integration, we have calculated the {\bf on-axis} 3-$\sigma$ sensitivity of LECX for 1, 10, 100 and 1000 seconds. The results are shown on Figure \ref{fig:sens}.

One can see that LECX is capable of detecting a wide range of typical GRB fluxes even with one-second integrations. For longer-duration GRBs, even somewhat weak events could be detected. The Crab spectrum \citep{2001ApJ...559..342W} is shown in the figure just for comparison purposes, since the Crab it is a strong standard candle in these energies. LECX will not be able to detect the Crab since it would require pointing and very long stabilized observations, capabilities that the nanoMIRAX satellite will not have.  
 
\subsection{Simulations of cosmic explosion detections}\label{sec:grb}

In order to predict the performance of LECX for observations of GRBs and related phenomena, we have carried out a series of Monte Carlo simulations taking into account typical GRB spectra and the predicted LECX background. {\bf Since the LECX's graded shield, which is the same as the one used in our balloon project protoMIRAX \citep{Braga2015}, has been shown by \citet{2016MNRAS.459.3917C} to be very effective in absorbing most gamma-rays in the typical GRB energy ranges, and also in cutting off the lead and tin fluorescence, we do not expect to have any significant contribution of scattering and transmission of the burst photons in the collimator.  We also expect that contamination from GRB photons scattering off the spacecraft structure will be negligible since there is very little material in the cubesat platform and the detector plane is well shielded from below. For the same reason, and also due to the fact that the background is highly dominated by the diffuse gamma-ray emission coming through the aperture up to $\sim$\ts 140\ts keV, as seen in section \ref{sec:sens}, the albedo contribution from the GRB photons themselves scattering off Earth's atmosphere is very likely to be unimportant. In these simulations, we did not take into account Compton scattered photons in one detector hitting another detector since these scatterings will be very rare (see below) and the geometrical cross section of one detector seen by another is extremely small. Taking these considerations into account, we found very reasonable, for estimation purposes, to carry our simulations in which the GRB photons only hit the detector surfaces.} 

GRBs are brief flashes of $\gamma$-rays with spectral energy distributions that peak around hundreds of keV (see \citealp{2018pgrb.book.....Z} for a recent review), {\bf which are expected to be detectable at a rate of $\sim$\ts 1 event per day in the entire sky with the currently available instrumentation \citep{2018ApJ...858...79W}}. They represent the most energetic explosions in the Universe and can release up to $10^{54}$ ergs in a few seconds. They follow a bimodal distribution in which most of the bursts last longer than $\sim$\ts 2 seconds, clustering around tens of seconds, and about $1/ 3$ of them are shorter, clustering around 400\ts ms. The former are believed to be produced by the core collapse of massive, high-rotation stars, whereas the latter are best explained by the coalescence of neutron stars in a binary system ({\bf e.g.}\ \citealp{2014ARA&A..52...43B}). These {\bf short bursts} are additionally interesting because they are expected to emit copious amounts of energy in the form of gravitational waves, as was dramatically demonstrated by the GW\ts 170817 event ({\bf e.g.}\ \citealp{2017ApJ...850L..24G, 2017ApJ...848L..12A, 2017ApJ...848L..15S}). As the current gravitational wave detectors (LIGO and VIRGO) are upgrading their sensitivities and new ones (KAGRA, LIGO-India) are about to start operating, it is expected that the rate of GW burst discoveries will significantly increase over the next years. It is extremely important that those detections are accompanied simultaneously by X- and $\gamma$-ray observations, since the GW instruments lack precise angular localization and the EM signals are complementary with respect to the characterisation of the source. In a sense, we ``see'' the event through EM signals and ``hear'' it through the GW signal.

Short GRBs (SGRBs) usually have harder spectra and are less energetic than long GRBs \citep{2014ARA&A..52...43B}. However, the fluxes detected {\bf at Earth} vary by several orders of magnitude depending on the GRB distance. The time-integrated fluxes (fluences) for all GRBs range {\bf approximately} from $10^{-8}$ to $10^{-4}$ erg cm$^{-2}$. The observed photon spectra also show significant diversity and can usually be fit with the so-called Band model, a broken power-law with a smooth junction \citep{1993ApJ...413..281B}. In any event, since we are interested in order-of-magnitude values for the simulations, we can approximately consider that a typical GRB photon spectrum measured at Earth is $ F = A E^{-1}$, where $A$ varies from $0.25$ (weakest GRBs) to 200 (strongest GRBs), $E$ is measured in keV and $F$ is given in photons \fl. {\bf This corresponds to a flat $\nu F_\nu$ spectrum.} Figure \ref{fig:sens} shows the limiting cases of these spectra.

In the simulations, we have considered the LECX instrument in a near-equatorial LEO and an incident flux coming from a GRB in a given direction. If on-axis, a very strong GRB would produce $\sim$\ts 1,900\ts counts in 1\ts s in the 30-200\ts keV energy range, whereas a very weak GRB would produce $\sim$\ts 2\ts counts. The background rate calculated {\bf using the GEANT4 simulations} is $\sim$\ts 2.5\ts counts/s.

Using these numbers, we can simulate the signal-to-noise ratio (SNR) and the localization accuracy that we could achieve in a GRB detection with LECX. As in the simulations reported in section \ref{sec:posit}, we first calculate the illuminated areas on the four detectors, defined by the direction in the sky from which the photons are coming. In the real observations, this will depend not only on the azimuth and elevation with respect to the experiment {\bf reference frame} but also on the satellite attitude given by the spacecraft's attitude control system (see section \ref{sec:sat}). By combining the two coordinate systems, we will determine the celestial coordinates (e.g.\ right ascension and declination) of the event in the sky. 

Once the illuminated areas are defined, we simulate random source counts at the level of the known sources in each detector using a Poisson distribution. These numbers scale with the illuminated fraction of the detector areas. The source counts are calculated convolving the source spectra with the response function of the instrument, which is given by the ratio of the effective area to the geometrical area of the detectors as a function of energy. In this case this will be essentially the photoelectric efficiency of the CZT detectors, {\bf since in these detectors the photoelectric effect is highly favoured with respect to Compton scatterings up to several hundred keV \citep{2009sensors}}. The background counts added to that are also Poissonian distributed and are given by the values provided by the GEANT4 simulations.

\begin{figure}
\centering
\includegraphics[width=0.4\hsize]{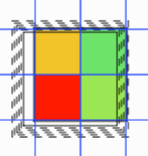}
\includegraphics[width=0.12\hsize]{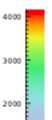}
\caption{The detector plane and shadow geometry for a very strong GRB simulation. The simulated position in the sky is $A = 23$\gr\ and $z = 7$\gr. The colour scale gives the count level at each detector.}
\label{fig:GRB_23_7}
\end{figure}

As a first {\bf somewhat unrealistic} example {\bf to demonstrate the method}, we simulated a 10-s very strong GRB, {\bf corresponding to a fluence of 5.4$\times 10^{-4}$ erg cm$ ^{-2}$ in our 30--200\ts keV range},  occurring at $A = 23$\gr\ and $z = 7$\gr. Figure \ref{fig:GRB_23_7} shows the count levels and the shadow position. 

\begin{figure}
\centering
\includegraphics[width=0.6\hsize]{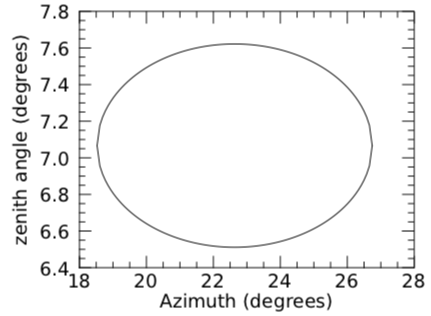}
\caption{A simulated localization for a strong GRB. The position algorithm gives $A = (23\pm4)\!$\gr\ and $z = (7.0\pm0.7)\!$\gr\ with respect to the experiment coordinates. This can be readily converted to sky coordinates based on the satellite attitude.}
\label{fig:pos_strong_GRB}
\end{figure}

In this case we get $\sim$\ts 13,600 source counts and $\sim$\ts 117 background counts, providing a theoretical (Poissonian) SNR of $\sim$\ts 116. After repeating the simulation 100 times to determine the dispersion on the parameters, we get $A = 23\pm4$\gr\ and $z = 7.0\pm0.7$\gr. The values of the dispersions do not vary significantly for more than $\sim$\ts 5 repetitions of the simulations, so these 1-$\sigma$ dispersion values are very robust. This GRB is detected at a level of 116$\sigma$ (SNR = 116.4$\pm$0.5), in close agreement with the Poissonian value. Figure \ref{fig:pos_strong_GRB} shows the 1-$\sigma$ sky localization region for the GRB. 

In a second example, {\bf more realistic for a short GRB}, we simulated a SGRB 10 times weaker with a duration of 1\ts s. {\bf The fluence in this case is 5.4$\times 10^{-6}$ erg cm$^{-2}$}. The incoming direction is $A = 203$\gr\ and $z=14$\gr. We get $\sim$\ts 120 source counts and $\sim$\ts 11 background counts, providing a theoretical (Poissonian) SNR of $\sim$\ts 11. The 100 simulations provide $A = 200\pm20$\gr\ and $z = 15.0\pm3.3$\gr, with a SNR of $10.4\pm0.5$, again in good agreement with the theoretical value.

It is noteworthy that the algorithm also gives the source flux during the event. By determining the source direction, we can calculate the fraction of the total detector area that was illuminated. Knowing the zenith angle $z$, we can correct for the the projected area by dividing the geometrical area by $\cos z$. Then, a deconvolution of the resulting flux in counts \fl\ by the instrument's response function gives the source's flux in photons \fl. 

{\bf Since the satellite attitude determination precision is of $\sim$\ts 1\gr\ and we do not expect the spacecraft to rotate by more than 1\gr\ in 10\ts s (see section \ref{sec:sat}), the conversion of instrument to sky coordinates will have an error of $\lesssim$\ts 1\gr, which will be a small fraction of the overall localization power for most detected GRBs.}

These simulations demonstrate that LECX is able to detect cosmic explosions and determine their positions in the sky with enough accuracy to contribute to the search for EM counterparts of GW burst events detected by the new generation of GW laser-interferometric observatories. 

{\bf Based on current estimates of GRB occurrence rates and LECX's sensitivity, we can estimate the rate at which we will be able to detect these events in our FOV. The GRB rates have been discussed extensively in the literature (e.g.\ \citealp{2016A&A...594A..84G, 2015MNRAS.448.2514G, 2018ApJ...858...79W}) and the values vary by large factors, depending on different models used for the GRB luminosity function and redshift distribution and on the assumptions made. Since the GRB emission is generally supposed to be collimated in jets, here we are counting only {\it detectable\/} GRBs, meaning that real GRB occurrence rate in the Universe can be much higher, depending on the jet opening angles. In order to get an estimate, we have used Figure 3 of \citet{2015MNRAS.448.2514G} that shows cumulative GRB rates in number-of-GRBs year$^{-1}$sr$^{-1}$ as a function of the peak flux in photons cm$^{-2}$ s$^{-1}$. Considering that the detected GRBs will have a roughly uniform angular distribution with respect to the instrument axis, we first apply a factor of $\sim$\ts 2 decrease in the detector plane illuminated area, corresponding to a typical $z$ angle of 25\gr. This will produce an increase in the minimum detectable flux (the sensitivity) by a factor of $\sim\!\!\sqrt{2}$. In order to use the 50-300\ts keV curve of Figure 3 of \citet{2015MNRAS.448.2514G}, which is the more adequate one to use here, we have to estimate the count rate that LECX would produce in this energy range. We used our GEANT4 simulations for the 50-200\ts keV range, providing 1.4 counts s$^{-1}$, and the fact that LECX would detect a very small number of events from 200 to 300\ts keV due to the relatively low efficiency of our 2-mm thick detector in this range and the low background (this is estimated to have an upper limit of 0.2 counts s$^{-1}$). Our estimation for the 10-s flux limit then gives $\sim$\ts 0.9 ph cm$^{-2}$s$^{-1}$, which corresponds to $\sim$\ts 11 GRBs yr$^{-1}$ sr$^{-1}$ in the above-mentioned figure (for all redshifts). Since our FWHM FOV covers 7\% of the sky, this would allow us to detect $\sim$\ts 10 events per year or $\sim$\ts 1 event every 36 days. Working with a conservative estimate that the instrument will be pointed away from Earth for half the time (see section \ref{sec:sat}), the rate of detected/localized events would be a more realistic $\sim$\ts 5 events per year.

It is important to stress that the GRB rate in the Universe is still subject to considerable debate, and the fact that the only localized neutron-star/neutron star coalescence event localized so far (GW170817 -- \citealp{2017ApJ...848L..12A}) was also seen in gamma rays \citep{2017ApJ...848L..13A} seems to point towards more frequent multi-messenger observations of these events than previously thought.}  

\section{The nanoMIRAX satellite} 
\label{sec:sat}

The LECX experiment is the payload of a 2U cubesat-based nanosatellite called nanoMIRAX or CRON-1 \citep{Durao2019}. The upper ``U'' comprises essentially the 3 PCBs that constitute the LECX experiment as described in section \ref{sec:det}. The lower ``U'' is the service module, which includes 4 subsystems: the Attitude Determination and Control Module (ADCM), the Power Generation and Management Module (PGGM), the On-Board Computer (OBC), and the Communications Module (COMM).

The ADCM, developed by Horuseye Tech, has been designed with potential for pointing at a target. However, due to cost limitations and the availability of sensors, this spacecraft only provides stabilization of angular velocity after release of the cubesat, using magnetic torque actuators, and attitude determination with magnetometers, solar sensors and gyrometers. The ADCM has an innovative design with redundant sensors for increased reliability. After release, the cubesat will be stabilized using the measurements of a triad of MEMS gyros, which provide the angular velocity of the cubesat with respect to an inertial frame. The actuation will be done by magnetic torquers, which generate orthogonal proportional torques at the inertial angular velocity and magnetic field, expressed in the cubesat axes. {\bf It is expected that the actuation will be capable of keeping the LECX's FOV away from the Earth for most of the time.} The attitude determination phase starts after stabilization in angular speed. The attitude is determined using the measurements of a 3-axis magnetometer and the solar sensors in the cubesat panels when there is no obscuration of the sun. It is expected that the system will provide attitude knowledge with a $\sim$\ts 1\gr\ average accuracy and an angular velocity limit of $\sim$\ts 0\gr $\!\!.1$\!/s. The ADCM software will store the attitude information during an event and then process it to reach the best possible accuracy taking into account the precision of the onboard sensors and the limits of magnetic actuation.

The PGGM, also developed by Horuseye Tech, will optimise the energy transfer from the solar panels by converting the power with BOOST-type converters using MPPT (Maximum Power Point Tracking) mode. The MPPT operation seeks to maximize the utilization of the energy that the panels can generate. This is useful when the system needs to recharge the batteries and does not consume the full capacity of the panels. In this case, all excess consumption is directed to storage. The implemented circuit can operate in two MPPT modes: constant voltage and maximum power. The PGGM uses a high-performance, low-power Atmel microcontroller (IC1), ARM 32bits, M4, of the SAMG55 family. The system is expected to provide 800mW of power to LECX continuously, which will allow the experiment to be on all the time, maximizing the search for cosmic explosions.

The OBC, acquired from ISIS (Innovative Solutions in Space), uses a 400MHz 32-bit ARM9 processor. The data communications protocol is I2C with 400 kbps maximum capacity. The OBC will receive science and housekeeping data from the LECX experiment and store them in data files to be downloaded to the ground stations.

The COMM will use a TRXUV CubeSat transceiver also provided by ISIS. It operates in the 130-160\ts MHz (VHF) and 482-486 MHz (UHF) frequency ranges with a data rate of 1200--9600 bit/s. We will use small-satellite ground stations already operational at INPE (National Institute for Space Research in Brazil) at the Brazilian cities of Santa Maria, S\~ao Jos\'e dos Campos and Natal.

{\bf All data acquired in the spacecraft will be transmitted to the ground and stored at the mission center. These include the science, attitude and housekeeping data. The onboard GPS will date all events with Universal Time. The science data will consist in all detected events in the 4 detectors with energy and time information (see section \ref{sec:det}). The attitude data will include the satellite attitude and stabilization information as a function of time. The housekeeping data will include temperatures in several locations within the spacecraft and other diverse ``health'' parameters of the satellite.}

In Table \ref{tab:sat} we present the baseline numbers of the LECX/nanoMIRAX mission.

\begin{table}
\centering
\caption{nanoMIRAX/LECX baseline parameters}
\label{tab:sat}
\begin{tabular}{l}
\textbf{LECX payload}\\
\hline
Detector type: CdZnTe (CZT)  \\
\hspace{0.3cm} Dimensions:  10\thinspace mm $\times$ 10\thinspace mm $\times$ 2\thinspace mm (thickness)\\
\hspace{0.3cm} Number of detectors:  4 ($2 \times 2$) \\
\hspace{0.3cm} Gap between detectors:  $3$ mm\\
Energy Range:  $30 - 200$ keV \\
Geometrical area:  $4$ cm$^2$\\
Effective area: $3.9$ cm$^2$@ 80 keV, $2.1$ cm$^2$@ 150 keV  \\
Energy resolution: 11\% @ 60 keV \\
Time resolution:  255 $\mu$s\\
Expected nominal counting rate: 2.5 counts/s\\ 
\hspace{0.3cm} (up to $\sim$\ts 100 counts/s in burst mode)\\
Shielding:  Pb (external -- 1.0\thinspace mm) \\
\hspace{1.3cm} Sn (middle -- 1.7\thinspace mm) \\
\hspace{1.3cm} Cu (internal -- 0.3\thinspace mm) \\
\hspace{1.3cm} configuration: box around detectors with  \\
\hspace{1.4cm} 23mm $\times$ 23mm aperture on top\\
Field of view: $53^{\circ} \times 53^{\circ}$ (FWHM); $90^{\circ} \times 90^{\circ}$ (FWZI)\\
Sensitivity: $10^{-2}$photons cm$^{-2}$s$^{-1}$keV$^{-1}$ for 1\ts s @ 100\ts keV\\
Cosmic Explosion (CE) Location Accuracy: a few degrees \\
\hspace{0.3cm} (depending on position and intensity)  \\
Expected CE detection rate: $\sim$\ts 5 per year\\
Science data creation rate: 100 bits/s\\
Payload mass: $\sim$\ts 600\ts g\\
\hline
{\textbf{nanoMIRAX satellite}}\\
\hline
Structure: 2U cubesat frame\\
Satellite dimensions: 20\thinspace cm $\times$ 10\thinspace cm $\times$ 10\thinspace cm \\
Total mass:  $\sim$\ts 2.1\ts kg (including payload)\\
Power: 2.2 W total, 800\ts mW delivered to payload\\
Stabilization: $\lesssim$ 0\gr$\!\!.1$/s\\
Attitude knowledge: $\sim$\ts 1\gr\\
Communications: Telemetry (download): UHF, 482--486\ts MHz\\
\hspace{1.3cm}  Command upload: VHF, 130--160\ts MHz  \\
\hspace{1.3cm} Both at 1200--9600 bit/s\\
Orbit: TBD (possibly polar LEO if launched by PSLV) \\
Ground stations: 3 in Brazil\\
\hline
\end{tabular}
\end{table}

\section{Conclusions}\label{sec:conc}

In this paper we present a very small high-energy astrophysics space experiment that is capable of contributing to the detection and localization of cosmic explosions that reveal themselves by copious emission of hard X-rays and low-energy $\gamma$-rays. The nanosat/cubesat revolution has opened new opportunities for the development of scientific space missions in the ``smaller, faster and cheaper'' paradigm. With innovative ideas, it is possible to contribute to science with low budgets and limited resources. The experiment described here is an example of a cubesat mission that is capable of doing competitive and important science by detecting and localizing cosmic explosions in the gravitational wave era of astronomy. This is accomplished by a new algorithm that uses the shadowing of shielding walls over the detector plane to reconstruct the incoming directions of photons coming from a bright and short-lived cosmic point source. 

Simulations described here show that LECX can detect most GRBs and locate them within a few degrees in the sky at a rate of {\bf $\sim$\ts 5 per year.} With a constellation of satellites of this kind, it would be possible to increase this rate significantly and provide a low-cost, fast-development network of electromagnetic localizers of gravitational wave events that could complement the observations made by the ground-based gravitational wave observatories. {\bf The development of such a constellation is one of our long-term goals and will be addressed elsewhere.}

LECX is currently in the integration and testing phase of its engineering model. The flight model will integrate the solar panels and onboard antennas. We are analyzing launch opportunities and we expect to have the satellite launched by the end of 2020. One possibility is a launch in a LEO polar orbit by the Indian Polar Satellite Launch Vehicle (PSLV).

\section*{Acknowledgements}

We thank FAPESP, Brazil, for support under Projeto Tem\'atico 2013/26258-4 and PIPE project 2017/09800-0. 
P.\ E.\ Stecchini acknowledges FAPESP PhD fellowship 2017/13551-6. 




\bibliography{refs}

\begin{thebibliography}{}
\makeatletter
\relax
\def\mn@urlcharsother{\let\do\@makeother \do\$\do\&\do\#\do\^\do\_\do\%\do\~}
\def\mn@doi{\begingroup\mn@urlcharsother \@ifnextchar [ {\mn@doi@}
  {\mn@doi@[]}}
\def\mn@doi@[#1]#2{\def\@tempa{#1}\ifx\@tempa\@empty \href
  {http://dx.doi.org/#2} {doi:#2}\else \href {http://dx.doi.org/#2} {#1}\fi
  \endgroup}
\def\mn@eprint#1#2{\mn@eprint@#1:#2::\@nil}
\def\mn@eprint@arXiv#1{\href {http://arxiv.org/abs/#1} {{\tt arXiv:#1}}}
\def\mn@eprint@dblp#1{\href {http://dblp.uni-trier.de/rec/bibtex/#1.xml}
  {dblp:#1}}
\def\mn@eprint@#1:#2:#3:#4\@nil{\def\@tempa {#1}\def\@tempb {#2}\def\@tempc
  {#3}\ifx \@tempc \@empty \let \@tempc \@tempb \let \@tempb \@tempa \fi \ifx
  \@tempb \@empty \def\@tempb {arXiv}\fi \@ifundefined
  {mn@eprint@\@tempb}{\@tempb:\@tempc}{\expandafter \expandafter \csname
  mn@eprint@\@tempb\endcsname \expandafter{\@tempc}}}

\bibitem[\protect\citeauthoryear{{Abbott} et~al.,}{{Abbott}
  et~al.}{2017a}]{2017ApJ...848L..12A}
{Abbott} B.~P.,  et~al., 2017a, \mn@doi [\apjl] {10.3847/2041-8213/aa91c9},
  \href {https://ui.adsabs.harvard.edu/abs/2017ApJ...848L..12A} {848, L12}

\bibitem[\protect\citeauthoryear{{Abbott} et~al.,}{{Abbott}
  et~al.}{2017b}]{2017ApJ...848L..13A}
{Abbott} B.~P.,  et~al., 2017b, \mn@doi [\apjl] {10.3847/2041-8213/aa920c},
  \href {https://ui.adsabs.harvard.edu/abs/2017ApJ...848L..13A} {848, L13}

\bibitem[\protect\citeauthoryear{Agostinelli et~al.,}{Agostinelli
  et~al.}{2003}]{Agostinelli2003}
Agostinelli S.,  et~al., 2003, \mn@doi [Nuclear Instruments and Methods in
  Physics Research Section A: Accelerators, Spectrometers, Detectors and
  Associated Equipment] {http://dx.doi.org/10.1016/S0168-9002(03)01368-8}, 506,
  250

\bibitem[\protect\citeauthoryear{{Ajello} et~al.,}{{Ajello}
  et~al.}{2008}]{Ajello2008}
{Ajello} M.,  et~al., 2008, \mn@doi [The Astrophysical Journal]
  {10.1086/592595}, \href {http://adsabs.harvard.edu/abs/2008ApJ...689..666A}
  {689, 666}

\bibitem[\protect\citeauthoryear{{Armstrong}, {Chandler}  \&
  {Barish}}{{Armstrong} et~al.}{1973}]{Armstrong1973}
{Armstrong} T.~W.,  {Chandler} K.~C.,   {Barish} J.,  1973, \mn@doi [Jornal of
  Geophysical Research] {10.1029/JA078i016p02715}, \href
  {http://adsabs.harvard.edu/abs/1973JGR....78.2715A} {78, 2715}

\bibitem[\protect\citeauthoryear{{Band} et~al.,}{{Band}
  et~al.}{1993}]{1993ApJ...413..281B}
{Band} D.,  et~al., 1993, \mn@doi [\apj] {10.1086/172995}, \href
  {https://ui.adsabs.harvard.edu/abs/1993ApJ...413..281B} {413, 281}

\bibitem[\protect\citeauthoryear{{Berger}}{{Berger}}{2014}]{2014ARA&A..52...43B}
{Berger} E.,  2014, \mn@doi [\araa] {10.1146/annurev-astro-081913-035926},
  \href {https://ui.adsabs.harvard.edu/abs/2014ARA&A..52...43B} {52, 43}

\bibitem[\protect\citeauthoryear{{Braga}}{{Braga}}{2006}]{2006AIPC..840....3B}
{Braga} J.,  2006, in {D'Amico} F.,  {Braga} J.,   {Rothschild} R.~E.,  eds,
  American Institute of Physics Conference Series Vol. 840, The Transient Milky
  Way: A Perspective for MIRAX. pp~3--7, \mn@doi{10.1063/1.2216594}

\bibitem[\protect\citeauthoryear{{Braga}}{{Braga}}{2020}]{2020PASP..132a2001B}
{Braga} J.,  2020, \mn@doi [\pasp] {10.1088/1538-3873/ab450a}, \href
  {https://ui.adsabs.harvard.edu/abs/2020PASP..132a2001B} {132, 012001}

\bibitem[\protect\citeauthoryear{{Braga} et~al.,}{{Braga}
  et~al.}{2004}]{2004AdSpR..34.2657B}
{Braga} J.,  et~al., 2004, \mn@doi [Advances in Space Research]
  {10.1016/j.asr.2003.03.062}, \href
  {http://adsabs.harvard.edu/abs/2004AdSpR..34.2657B} {34, 2657}

\bibitem[\protect\citeauthoryear{{Braga} et~al.,}{{Braga}
  et~al.}{2015}]{Braga2015}
{Braga} J.,  et~al., 2015, \mn@doi [\aap] {10.1051/0004-6361/201526343}, \href
  {http://adsabs.harvard.edu/abs/2015A%26A...580A.108B} {580, A108}

\bibitem[\protect\citeauthoryear{{Castro}, {Braga}, {Penacchioni}, {D'Amico}
  \& {Sacahui}}{{Castro} et~al.}{2016}]{2016MNRAS.459.3917C}
{Castro} M.,  {Braga} J.,  {Penacchioni} A.,  {D'Amico} F.,   {Sacahui} R.,
  2016, \mn@doi [\mnras] {10.1093/mnras/stw743}, \href
  {https://ui.adsabs.harvard.edu/abs/2016MNRAS.459.3917C} {459, 3917}

\bibitem[\protect\citeauthoryear{{Dean}, {Bird}, {Diallo}, {Ferguson},
  {Lockley}, {Shaw}, {Westmore}  \& {Willis}}{{Dean} et~al.}{2003}]{Dean2003}
{Dean} A.~J.,  {Bird} A.~J.,  {Diallo} N.,  {Ferguson} C.,  {Lockley} J.~J.,
  {Shaw} S.~E.,  {Westmore} M.~J.,   {Willis} D.~R.,  2003, \mn@doi [Space
  Science Reviews] {10.1023/A:1023995803108}, \href
  {http://adsabs.harvard.edu/abs/2003SSRv..105..285D} {105, 285}

\bibitem[\protect\citeauthoryear{{Del Sordo}, {Abbene}, {Caroli}, {Mancini},
  {Zappettini}  \& {Ubertini}}{{Del Sordo} et~al.}{2009}]{2009sensors}
{Del Sordo} S.,  {Abbene} L.,  {Caroli} E.,  {Mancini} A.~M.,  {Zappettini} A.,
    {Ubertini} P.,  2009, \mn@doi [Sensors] {10.3390/s90503491}, 9, 3491

\bibitem[\protect\citeauthoryear{{Dur\~ao}, {Braga}, {Schad}, {Silva Neto},
  {Esper}, {Rigobello}  \& {Ribeiro}}{{Dur\~ao} et~al.}{2019}]{Durao2019}
{Dur\~ao} O.~S.~C.,  {Braga} J.,  {Schad} V.~R.,  {Silva Neto} M.~B.,  {Esper}
  M.,  {Rigobello} G.,   {Ribeiro} C.~B.,  2019, Proceedings of the 2019 Small
  Satellite Conference, p.~9

\bibitem[\protect\citeauthoryear{{Fenimore} \& {Cannon}}{{Fenimore} \&
  {Cannon}}{1978}]{1978ApOpt..17..337F}
{Fenimore} E.~E.,  {Cannon} T.~M.,  1978, \mn@doi [\ao] {10.1364/AO.17.000337},
  \href {http://adsabs.harvard.edu/abs/1978ApOpt..17..337F} {17, 337}

\bibitem[\protect\citeauthoryear{{Gehrels}}{{Gehrels}}{1985}]{Gehrels1985}
{Gehrels} N.,  1985, \mn@doi [Nuclear Instruments and Methods in Physics
  Research A] {10.1016/0168-9002(85)90732-6}, \href
  {http://adsabs.harvard.edu/abs/1985NIMPA.239..324G} {239, 324}

\bibitem[\protect\citeauthoryear{{Ghirlanda} et~al.,}{{Ghirlanda}
  et~al.}{2015}]{2015MNRAS.448.2514G}
{Ghirlanda} G.,  et~al., 2015, \mn@doi [\mnras] {10.1093/mnras/stv183}, \href
  {https://ui.adsabs.harvard.edu/abs/2015MNRAS.448.2514G} {448, 2514}

\bibitem[\protect\citeauthoryear{{Ghirlanda} et~al.,}{{Ghirlanda}
  et~al.}{2016}]{2016A&A...594A..84G}
{Ghirlanda} G.,  et~al., 2016, \mn@doi [\aap] {10.1051/0004-6361/201628993},
  \href {https://ui.adsabs.harvard.edu/abs/2016A&A...594A..84G} {594, A84}

\bibitem[\protect\citeauthoryear{{Granot}, {Guetta}  \& {Gill}}{{Granot}
  et~al.}{2017}]{2017ApJ...850L..24G}
{Granot} J.,  {Guetta} D.,   {Gill} R.,  2017, \mn@doi [\apjl]
  {10.3847/2041-8213/aa991d}, \href
  {https://ui.adsabs.harvard.edu/abs/2017ApJ...850L..24G} {850, L24}

\bibitem[\protect\citeauthoryear{{Gruber}, {Matteson}, {Peterson}  \&
  {Jung}}{{Gruber} et~al.}{1999}]{Gruber1999}
{Gruber} D.~E.,  {Matteson} J.~L.,  {Peterson} L.~E.,   {Jung} G.~V.,  1999,
  \mn@doi [The Astrophysical Journal] {10.1086/307450}, \href
  {http://adsabs.harvard.edu/abs/1999ApJ...520..124G} {520, 124}

\bibitem[\protect\citeauthoryear{{Pike} et~al.,}{{Pike}
  et~al.}{2018}]{2018SPIE10762E..14P}
{Pike} S.~N.,  et~al., 2018, in \procspie. p. 1076214,
  \mn@doi{10.1117/12.2321990}

\bibitem[\protect\citeauthoryear{{Rodrigues}, {Grindlay}, {Allen}, {Hong},
  {Barthelmy}, {Braga}, {D'Amico}  \& {Rothschild}}{{Rodrigues}
  et~al.}{2013}]{2013JInst...8P9010R}
{Rodrigues} B.~H.~G.,  {Grindlay} J.~E.,  {Allen} B.,  {Hong} J.,  {Barthelmy}
  S.,  {Braga} J.,  {D'Amico} F.,   {Rothschild} R.~E.,  2013, \mn@doi [Journal
  of Instrumentation] {10.1088/1748-0221/8/09/P09010}, \href
  {https://ui.adsabs.harvard.edu/abs/2013JInst...8P9010R} {8, P09010}

\bibitem[\protect\citeauthoryear{{Sarkar}, {Mandal}, {Debnath}, {Kotoch},
  {Nandi}, {Rao}  \& {Chakrabarti}}{{Sarkar} et~al.}{2011}]{Sarkar2011}
{Sarkar} R.,  {Mandal} S.,  {Debnath} D.,  {Kotoch} T.~B.,  {Nandi} A.,  {Rao}
  A.~R.,   {Chakrabarti} S.~K.,  2011, \mn@doi [Experimental Astronomy]
  {10.1007/s10686-010-9208-z}, \href
  {http://adsabs.harvard.edu/abs/2011ExA....29...85S} {29, 85}

\bibitem[\protect\citeauthoryear{{Savchenko} et~al.,}{{Savchenko}
  et~al.}{2017}]{2017ApJ...848L..15S}
{Savchenko} V.,  et~al., 2017, \mn@doi [\apjl] {10.3847/2041-8213/aa8f94},
  \href {https://ui.adsabs.harvard.edu/abs/2017ApJ...848L..15S} {848, L15}

\bibitem[\protect\citeauthoryear{{Villela}, {Costa}, {Brand\~ao}, {Bueno}  \&
  {Leonardi}}{{Villela} et~al.}{2019}]{Thyrso2019}
{Villela} T.,  {Costa} C.~A.,  {Brand\~ao} A.~M.,  {Bueno} F.~T.,   {Leonardi}
  R.,  2019, \mn@doi [International Journal of Aerospace Engineering]
  {10.1155/2019/5063145}, 2019, 13

\bibitem[\protect\citeauthoryear{{Wallyn}, {Ling}, {Mahoney}, {Wheaton}  \&
  {Durouchoux}}{{Wallyn} et~al.}{2001}]{2001ApJ...559..342W}
{Wallyn} P.,  {Ling} J.~C.,  {Mahoney} W.~A.,  {Wheaton} W.~A.,   {Durouchoux}
  P.,  2001, \mn@doi [\apj] {10.1086/322318}, \href
  {https://ui.adsabs.harvard.edu/abs/2001ApJ...559..342W} {559, 342}

\bibitem[\protect\citeauthoryear{{Williams}, {Clark}, {Williamson}  \&
  {Heng}}{{Williams} et~al.}{2018}]{2018ApJ...858...79W}
{Williams} D.,  {Clark} J.~A.,  {Williamson} A.~R.,   {Heng} I.~S.,  2018,
  \mn@doi [\apj] {10.3847/1538-4357/aab847}, \href
  {https://ui.adsabs.harvard.edu/abs/2018ApJ...858...79W} {858, 79}

\bibitem[\protect\citeauthoryear{{Zhang}}{{Zhang}}{2018}]{2018pgrb.book.....Z}
{Zhang} B.,  2018, {The Physics of Gamma-Ray Bursts}.
Cambridge University Press, \mn@doi{10.1017/9781139226530}

\makeatother
\end{thebibliography}
\bibliographystyle{mnras}



\appendix
%
%


\bsp	
\label{lastpage}
\end{document}